\documentclass[]{ANStopical}
\pdfoutput=1
\usepackage{booktabs}
\usepackage{bm}
\newcommand{\nSamples}{250}
\newcommand{\nSpline}{50}
\newcommand{\nCurved}{100}
\newcommand{\nManhattan}{100}
\newcommand{\nCapped}{189}
\newcommand{\nUncapped}{61}
\newcommand{\nBatchesGen}{200}
\newcommand{\nParticlesGen}{10,000}
\newcommand{\concreteMin}{2}
\newcommand{\concreteMax}{51}
\newcommand{\targetMinMedian}{-11}
\newcommand{\targetMinMin}{-45}
\newcommand{\nBelowMinusFifteen}{71}
\newcommand{\nUnscoredMaps}{2}
\newcommand{\unscoredCellsMax}{170}
\newcommand{\unscoredCellsMin}{5}
\newcommand{\nTargetPixels}{640,800}
\newcommand{\rtNodeHours}{165}

\newcommand{\rtAllMaxH}{6.2}

\newcommand{\rtUncMeanMin}{73}
\newcommand{\rtUncMedianMin}{57}

\newcommand{\rtUncMaxH}{6.2}

\newcommand{\rtCapMeanMin}{8}
\newcommand{\rtCapMedianMin}{7}

\newcommand{\rtCapMaxMin}{18}
\newcommand{\rtCapNodeH}{24}

\newcommand{\ccNGeoms}{4}
\newcommand{\ccSpeedMin}{15}
\newcommand{\ccSpeedMax}{184}
\newcommand{\ccSigmaBright}{0.003}
\newcommand{\ccSigmaMid}{0.02}
\newcommand{\ccSigmaTailMin}{0.3}
\newcommand{\ccSigmaTailMax}{1.4}
\newcommand{\xsTf}{0.348}
\newcommand{\xsTt}{0.540}
\newcommand{\xsSf}{0.348}
\newcommand{\xsSt}{0.532}
\newcommand{\xsAirTf}{1.39 \times 10^{-4}}
\newcommand{\xsAirTt}{5.14 \times 10^{-4}}
\newcommand{\xsAirSf}{1.35 \times 10^{-4}}
\newcommand{\xsAirSt}{4.56 \times 10^{-4}}
\newcommand{\baseRmseA}{3.16}
\newcommand{\baseTenA}{28}
\newcommand{\unetRmseA}{1.19}
\newcommand{\unetMaeA}{0.52}
\newcommand{\unetTenA}{89}

\newcommand{\nTrain}{200}
\newcommand{\nVal}{50}
\newcommand{\nParams}{1.9}
\newcommand{\baseRmseB}{2.89}
\newcommand{\baseTenB}{28}
\newcommand{\unetRmseB}{1.20}
\newcommand{\unetMaeB}{0.46}
\newcommand{\unetTenB}{90}

\newcommand{\baseRmseC}{3.06}
\newcommand{\baseTenC}{30}
\newcommand{\unetRmseC}{1.00}
\newcommand{\unetMaeC}{0.46}
\newcommand{\unetTenC}{90}

\newcommand{\unetRmseMin}{1.00}
\newcommand{\unetRmseMax}{1.20}

\newcommand{\trainRmseMin}{0.46}
\newcommand{\trainRmseMax}{0.56}

\newcommand{\lcMeanNTwentyFive}{2.15}
\newcommand{\lcMeanNFifty}{1.79}
\newcommand{\lcMeanNHundred}{1.62}
\newcommand{\lcMeanNTwoHundred}{1.13}
\newcommand{\stratPixels}{480,600}

\newcommand{\stratShareA}{0.3}
\newcommand{\stratRmseA}{0.21}
\newcommand{\stratTenA}{100}
\newcommand{\roughA}{0.20}
\newcommand{\stratShareB}{24}
\newcommand{\stratRmseB}{0.14}
\newcommand{\stratTenB}{100}
\newcommand{\roughB}{0.05}
\newcommand{\stratShareC}{31}
\newcommand{\stratRmseC}{0.35}
\newcommand{\stratTenC}{97}
\newcommand{\roughC}{0.09}
\newcommand{\stratShareD}{15}
\newcommand{\stratRmseD}{0.51}
\newcommand{\stratTenD}{95}
\newcommand{\roughD}{0.12}
\newcommand{\stratShareE}{20}
\newcommand{\stratRmseE}{0.98}
\newcommand{\stratTenE}{87}
\newcommand{\roughE}{0.27}
\newcommand{\stratShareF}{10}
\newcommand{\stratRmseF}{3.13}
\newcommand{\stratTenF}{38}
\newcommand{\roughF}{1.02}
\newcommand{\stratShareBright}{55}
\newcommand{\nfShallowAll}{0.01}
\newcommand{\nfMidAll}{0.03}
\newcommand{\nfDeepAll}{0.29}
\newcommand{\nfDeepBright}{0.006}
\newcommand{\nfDeepMinusFour}{0.027}
\newcommand{\nfDeepMinusSix}{0.22}
\newcommand{\nfDeepTail}{0.78}

\newcommand{\peakBias}{0.51}

\newcommand{\fitSlope}{1.07}
\newcommand{\fitIntercept}{0.09}
\newcommand{\peakBiasMaps}{150}
\newcommand{\peakBiasDucts}{125}

\newcommand{\peakBiasMin}{0.12}
\newcommand{\peakBiasMax}{0.82}
\newcommand{\predMsEnsemble}{15}
\newcommand{\predMsSingle}{5}

\newcommand{\caseACorr}{0.96}
\newcommand{\caseATen}{81}
\newcommand{\caseAOffset}{-0.45}
\newcommand{\caseACells}{1,044}
\newcommand{\caseARmseBright}{0.65}
\newcommand{\caseATenBright}{89}
\newcommand{\caseANDeep}{181}
\newcommand{\caseADeepMedianMagic}{-7.0}
\newcommand{\caseADeepMedianCnn}{-8.1}
\newcommand{\caseAMinCnn}{-17.8}
\newcommand{\caseAMinMagic}{-12.7}

\newcommand{\caseAEnsSdBright}{0.17}
\newcommand{\caseAEnsSdDeep}{1.26}

\newcommand{\fomAnRe}{0.0456}
\newcommand{\fomAnSpb}{0.35}
\newcommand{\fomAnHours}{0.10}
\newcommand{\fomAn}{1.4}

\newcommand{\fomGenRe}{0.00073}
\newcommand{\fomGenSpb}{42.3}
\newcommand{\fomGenHours}{11.8}
\newcommand{\fomGen}{44.4}
\newcommand{\fomGenRatio}{32}

\newcommand{\fomGenCapRe}{0.0062}
\newcommand{\fomGenCapSpb}{25.7}
\newcommand{\fomGenCapHours}{7.1}
\newcommand{\fomGenCap}{1.0}
\newcommand{\fomGenCapRatio}{0.7}

\newcommand{\fomMgRe}{0.00057}
\newcommand{\fomMgSpb}{32.3}
\newcommand{\fomMgHours}{9.0}
\newcommand{\fomMg}{93.9}
\newcommand{\fomMgRatio}{68}

\newcommand{\fomMgCapRe}{0.0083}
\newcommand{\fomMgCapSpb}{20.6}
\newcommand{\fomMgCapHours}{5.7}
\newcommand{\fomMgCap}{0.7}
\newcommand{\fomMgCapRatio}{0.5}

\newcommand{\fomFlFRe}{0.0012}
\newcommand{\fomFlFSpb}{12.3}
\newcommand{\fomFlFHours}{3.4}
\newcommand{\fomFlF}{53.8}
\newcommand{\fomFlFRatio}{39}

\newcommand{\fomFlTRe}{0.0011}
\newcommand{\fomFlTSpb}{15.5}
\newcommand{\fomFlTHours}{4.3}
\newcommand{\fomFlT}{57.7}
\newcommand{\fomFlTRatio}{42}

\newcommand{\fomClFRe}{0.00073}
\newcommand{\fomClFSpb}{26.3}
\newcommand{\fomClFHours}{7.3}
\newcommand{\fomClF}{71.4}
\newcommand{\fomClFRatio}{51}

\newcommand{\fomClTRe}{0.00065}
\newcommand{\fomClTSpb}{31.6}
\newcommand{\fomClTHours}{8.8}
\newcommand{\fomClT}{74.8}
\newcommand{\fomClTRatio}{54}

\newcommand{\fomNfFBatches}{50}

\newcommand{\clampOverReuse}{0.80}
\newcommand{\clampOverGen}{1.7}
\newcommand{\clampOverFloor}{1.3}

\newcommand{\clampReOverMagic}{1.13}
\newcommand{\clampSpbOverMagic}{0.98}

\newcommand{\clampFSpbOverFloor}{2.1}
\newcommand{\floorSpbOverMagic}{0.38}
\newcommand{\floorReOverMagic}{2.1}
\newcommand{\capGenReRatio}{8.5}
\newcommand{\capGenSpeed}{1.6}
\newcommand{\capReuseReRatio}{14}
\newcommand{\capReuseCostSaving}{36}
\newcommand{\meshGainClamp}{5}
\newcommand{\meshGainFloor}{7}

\newcommand{\nodeHoursOverGen}{14}
\newcommand{\ttOneAn}{2.0 h$^*$}
\newcommand{\ttOneGen}{39 min}
\newcommand{\ttOneMg}{6 min}
\newcommand{\ttOneClF}{5 min}
\newcommand{\ttOneClT}{6 min}

\newcommand{\ttOneCapRoute}{7.2 h}
\newcommand{\ttHalfAn}{8.0 h$^*$}
\newcommand{\ttHalfGen}{1.1 h}
\newcommand{\ttHalfMg}{13 min}
\newcommand{\ttHalfClF}{13 min}
\newcommand{\ttHalfClT}{11 min}

\newcommand{\ttHalfCapRoute}{7.3 h}
\newcommand{\ttFifthAn}{50.0 h$^*$}
\newcommand{\ttFifthGen}{3.0 h}
\newcommand{\ttFifthMg}{52 min}
\newcommand{\ttFifthClF}{58 min}
\newcommand{\ttFifthClT}{59 min}

\newcommand{\ttFifthCapRoute}{8.0 h}
\newcommand{\ttTenthAn}{199.9 h$^*$}
\newcommand{\ttTenthGen}{7.3 h}
\newcommand{\ttTenthMg}{3.2 h}
\newcommand{\ttTenthClF}{3.7 h}
\newcommand{\ttTenthClT}{4.0 h}

\newcommand{\ttTenthCapRoute}{10.4 h}

\newcommand{\ttTenthClampOverReuse}{1.22}
\newcommand{\ttTenthCapGenHours}{277}

\newcommand{\meshMgCorrMedian}{188}

\newcommand{\meshClFCorrMedian}{201}

\newcommand{\meshClTCorrMedian}{187}

\newcommand{\cpMgMin}{-12.7}
\newcommand{\cpMgProxy}{1.00}
\newcommand{\cpMgMeas}{1.00}
\newcommand{\cpMgTop}{0.1}
\newcommand{\cpFiftyMin}{-9.2}
\newcommand{\cpFiftyProxy}{2.33}
\newcommand{\cpFiftyMeas}{2.15}
\newcommand{\cpFiftyTop}{1.6}
\newcommand{\cpFlMin}{-5.6}
\newcommand{\cpFlProxy}{0.42}
\newcommand{\cpFlMeas}{0.38}
\newcommand{\cpFlTop}{0.4}
\newcommand{\cpCpMin}{-9.8}
\newcommand{\cpCpProxy}{7.07}
\newcommand{\cpCpMeas}{9.36}
\newcommand{\cpCpTop}{3.6}
\newcommand{\cpNfMin}{-17.1}
\newcommand{\cpNfProxy}{53}
\newcommand{\cpNfMeas}{42}
\newcommand{\cpNfTop}{61}
\newcommand{\cpClMin}{-5.4}
\newcommand{\cpClProxy}{0.74}
\newcommand{\cpClMeas}{0.81}
\newcommand{\cpClTop}{0.9}
\newcommand{\cpNfDeepCells}{6}
\newcommand{\cpNfTopDecades}{4.4}
\newcommand{\cpMaxDev}{32}

\title{Predicting Weight Windows for Duct-Streaming Shielding Geometries with a Convolutional Neural Network}

\addAuthor[jlmiao@psu.edu]{Jilang Miao}{1}
\addAuthor{Ansel Kneier}{1}
\addAuthor{Jonathan Porrazzo}{1}
\addAffiliation{1}{Ken and Mary Alice Lindquist Department of Nuclear Engineering, The Pennsylvania State University, University Park, PA 16802, USA}

\Abstract{%
Forward-flux-based weight windows give global variance reduction for duct-streaming shielding problems,
but each new geometry needs an iterative Monte Carlo calculation to build them.
We report preliminary results on a convolutional neural network (CNN) that
learns the map from a two-dimensional duct geometry and source position to the lower window bound of such an iteration,
here the MAGIC generator in OpenMC,
in milliseconds.
A set of \nSamples{} random concrete ducts of three curved and rectangular families on the Protvino labyrinth bounding box was run in OpenMC with a Cf-252 point source and MAGIC generation,
and a \nParams{}-million-parameter U-Net was trained on the logarithm of the lower bound.
On held-out ducts the error in the cells that carry the flux is \stratRmseB{} to \stratRmseD{} decades;
replicate MAGIC runs show this part of the map to be reproducible to a few hundredths of a decade,
so the residual is model error, while below $10^{-6}$ the reference itself is noise.
Applied without retraining to the rectangular Protvino labyrinth,
the prediction correlates with MAGIC at \caseACorr.
Used as weight windows there, with its deepest cells clamped at deployment,
it gives an exit-detector figure of merit \fomClTRatio{} times the analog value,
\clampOverGen{} times that of the MAGIC generation run for the same geometry,
and \clampOverReuse{} of that of reused MAGIC windows whose construction it does not pay for.
The run cost of any window set is shown to be set by its deepest cells and predictable from a population proxy.
}

\shortTitle{Weight-Window Prediction for Duct Streaming with a CNN}
\authorHead{J. Miao et al.}

\begin{document}

\section{Introduction}\label{sec:intro}

Weight windows are the most widely used variance-reduction device in Monte Carlo shielding calculations:
a particle whose weight falls below the lower bound of the window in its cell is rouletted,
and one whose weight exceeds the upper bound is split,
so that the particle population is steered toward the regions the windows mark as important~\cite{Booth2006GenesisWW}.
The quality of the result depends on how the window bounds are obtained,
and the methods in use differ mainly in what they take as the measure of importance.

For a single detector response the natural measure is the adjoint flux.
The CADIS method~\cite{WagnerHaghighat1998CADIS} computes it with a deterministic discrete-ordinates solver and derives consistent source biasing and weight windows from it;
ADVANTG~\cite{Mosher2013ADVANTG} automates this workflow for MCNP.
When the quantity of interest is spread over the problem, as the dose along an access corridor is,
a single adjoint function is not the right target, and two families of global methods exist.
FW-CADIS~\cite{WagnerPeplowMosher2014FWCADIS} keeps the deterministic adjoint but weights its source by the inverse of the forward response,
so that relative uncertainties become uniform over the region of interest.
Forward-based schemes instead set the window bounds from the forward flux itself,
flattening the particle population:
the automated windows of Ref.~\cite{CooperLarsen2001AutomatedWW} iterate on the forward scalar flux,
the Serpent~2 scheme~\cite{leppanen2019global} adds a self-adaptive window mesh,
and the MAGIC iteration compared with FW-CADIS for ITER~\cite{davis2011gvr_iter} sets the lower bound in each mesh cell proportional to the forward flux accumulated so far and refines it batch by batch.

All of these methods share one cost: they must be rerun for every new geometry.
The deterministic methods need a transport solve on a mesh of the geometry,
and the forward iterative methods need a sequence of Monte Carlo batches.
We use MAGIC as the reference generator because it is available on the fly in OpenMC~\cite{romano2015openmc,openmc_docs_magic_vr},
not because it is the best of these methods; any of them could serve as the training target.
Its price is a full Monte Carlo calculation per geometry,
which in the production runs of this study ranged from under a minute to \rtAllMaxH{} hours of active time depending on how much splitting the geometry induced.

Duct and labyrinth designs are iterated over many variants that differ only in wall placement and thickness while materials and source stay fixed,
so the map from geometry to importance can be treated as a function to be learned from examples.
We ask whether a CNN trained on random ducts reproduces the lower window bound of an unseen geometry closely enough to serve as its weight window,
and what the predicted windows then cost to run.
The work extends our study of Markov-chain windows on the Protvino labyrinth~\cite{Miao2025MCVR,HeinrichsGlesmann2024Protvino},
which fixes the domain, materials, and mesh used here.
This paper reports preliminary results: a dataset of \nSamples{} geometries of three families,
the network, its accuracy on held-out ducts against the noise of the reference itself,
a transfer to the rectangular benchmark,
a figure-of-merit test of the predicted windows in OpenMC against both a reused MAGIC file and the MAGIC generation run,
and a population argument that predicts the run cost of any window set from the map alone.
The benchmark is a single geometry,
so the figure-of-merit numbers establish feasibility rather than performance bounds.

\section{Method}\label{sec:method}

\subsection{Random Duct Geometries and MAGIC Reference}\label{sec:data}

Each sample is a two-dimensional pixel map of $89\times36$ cells at 20~cm,
i.e.\ the $1780\times720$~cm bounding box of Protvino cases~1 and~6~\cite{HeinrichsGlesmann2024Protvino},
with vacuum boundaries on all four sides and infinite extent in $z$.
Materials are the Protvino concrete and air compositions~\cite{HeinrichsGlesmann2024Protvino},
and the source is an isotropic Cf-252 Watt spectrum point source at the center of the source cell.
Three families of ducts were generated.
In the first,
\nSpline{} ducts follow a cubic spline from a random interior source point to a random exit on one of the four edges,
chosen with probability weighted by its distance from the source so that ducts span the domain;
a concrete band of half-width equal to the duct radius plus a wall thickness is stamped along the spline and an air channel of the duct radius is carved down its center,
giving a thin-walled concrete duct in air,
with the concrete cap at the source end removed so that the source sits in an open pocket.
The duct radius varies smoothly along the path between 2 and 4 pixels and the wall thickness is sampled between 80 and 160~cm.
The second family, \nCurved{} ducts,
uses the same construction with zero to three prescribed turns and walls of 80 to 320~cm,
up to twice as thick as before.
The third, \nManhattan{} ducts, has axis-aligned legs joined by one to four right-angle turns,
a constant radius, and walls of 80 to 240~cm, the style of the Protvino labyrinths.
Concrete covers \concreteMin{} to \concreteMax\% of the domain across the set.

Each geometry was run in OpenMC~0.15 as a lattice with 20~cm pitch and the MAGIC weight-window generator on the same $89\times36$ mesh,
\nParticlesGen{} particles per batch for \nBatchesGen{} batches,
so the windows are refined \nBatchesGen{} times.
The target is $\log_{10} w_{\rm low}$ from the resulting weight-window file, one energy group,
normalized by MAGIC to a maximum of 0.5.
Cells never scored by MAGIC are excluded through a validity mask;
they occur in \nUnscoredMaps{} of the \nSamples{} maps, \unscoredCellsMin{} and \unscoredCellsMax{} cells in the deepest shadows.
The deepest cell of a map lies at $10^{\targetMinMedian}$ in the median geometry and at $10^{\targetMinMin}$ in the deepest,
and \nBelowMinusFifteen{} of the \nSamples{} maps reach below $10^{-15}$.

The thick-walled families exposed a property of on-the-fly window generation that governs the cost of a dataset.
As MAGIC deepens the windows behind a thick wall,
a particle that penetrates it is split at every window boundary it crosses,
and with OpenMC's current default limit of $10^{7}$ splits per history the generation of the thickest-walled geometries did not finish in a 20~hour allocation.
Capping the splits per history at 1000, the OpenMC default before version 0.15, resolves this.
The cap was checked on \ccNGeoms{} of the first-family geometries by rerunning them capped:
the generation ran \ccSpeedMin{} to \ccSpeedMax{} times faster,
and the resulting maps agree with the uncapped ones to within \ccSigmaBright{} decades in every cell above $10^{-4}$ and \ccSigmaMid{} decades in $(10^{-6},10^{-4}]$,
with differences of \ccSigmaTailMin{} to \ccSigmaTailMax{} decades only below $10^{-6}$,
where \Cref{sec:val} shows the reference itself to be noise.
MAGIC sets each bound from the flux accumulated in that cell relative to the maximum over the mesh,
and the cap only changes how many tracks estimate the flux in the deepest cells, not its expectation,
since a history that is no longer split continues with its weight rather than being killed;
the map above $10^{-6}$ is therefore unchanged.
The \nCapped{} geometries of the second and third families were therefore generated with the cap and the \nUncapped{} others without.
The uncapped generations took \rtUncMeanMin{} minutes of active time on average (median \rtUncMedianMin,
maximum \rtUncMaxH{} hours) and the capped ones \rtCapMeanMin{} minutes (median \rtCapMedianMin,
maximum \rtCapMaxMin{} minutes), for \rtNodeHours{} node-hours in all,
of which the capped \nCapped{} account for \rtCapNodeH.

\subsection{Network, Inputs, and Training}\label{sec:net}

The network is a two-dimensional U-Net~\cite{Ronneberger2015UNet}: three pooling levels,
32 to 256 features, two $3\times3$ convolutions with group normalization per level,
skip connections, $\nParams\times10^{6}$ parameters, against \nTargetPixels{} target cells in a training split before augmentation.
The architecture is chosen because the window map couples scales:
the bound in a deep cell is set by the walls between it and the source anywhere in the domain,
which the pooling pyramid covers with a receptive field of the whole mesh,
while the skip connections keep the pixel-scale wall edges at which the bound drops;
being fully convolutional, the network also applies unchanged to the larger domain of \Cref{sec:caseA}.
A comparison against simpler CNN or fully connected baselines on this dataset is left for the full study.
Its seven input channels are the
binary geometry;
the optical thicknesses $\Sigma_{t,g}\Delta x$ and $\Sigma_{s,g}\Delta x$ for a fast (0.625~eV to 20~MeV) and a thermal group;
a Gaussian bump of width 1.5 pixels on the source cell;
and a mask of the native domain,
so that samples of different shapes can share a padded canvas with the loss restricted to real cells.
The two-group cross sections were pooled from the production runs by summing the raw reaction-rate and flux tallies of the material-domain multigroup library over the first \nSpline{} geometries before taking the ratio:
for concrete $\Sigma_t = \xsTf$, $\xsTt$~cm$^{-1}$ and $\Sigma_s = \xsSf$, $\xsSt$~cm$^{-1}$ (fast,
thermal), and for air $\Sigma_t = \xsAirTf$, $\xsAirTt$~cm$^{-1}$ and $\Sigma_s = \xsAirSf$,
$\xsAirSt$~cm$^{-1}$.
Repeating the pooling over all \nSamples{} geometries changed no value by more than 0.2\%,
so the split cap of \Cref{sec:data}, under which the other \nCapped{} geometries were generated,
does not bias the material flux and reaction-rate tallies,
and the thick-walled and rectangular families do not shift the flux spectrum that weights the group collapse.
At one pixel size these channels are a function of the geometry channel;
they are kept so the representation carries over when the pixel size varies.

The loss is the masked mean-squared error of the standardized target over every cell MAGIC scored,
minimized with Adam~\cite{KingmaBa2015Adam} (learning rate $10^{-3}$, cosine decay, 400 epochs,
batch size 8) with horizontal and vertical flips as augmentation,
valid because all boundaries are vacuum.
The \nSamples{} samples are split \nTrain/\nVal{} into training and validation for three random splits,
and the validation error is evaluated after every epoch and the weights of the epoch with the lowest value are kept,
the model-selection form of early stopping~\cite{Prechelt2012EarlyStopping};
training takes about ten minutes on eight CPU threads.
The validation split therefore also selects the checkpoint and no separate test partition is held out at this dataset size,
so the held-out numbers of \Cref{sec:val} are validation numbers and only \Cref{sec:caseA} is data that played no role in training.
Two variants of the loss were also trained on the same splits,
since \Cref{sec:val} shows that cells below $10^{-6}$ are noise as targets:
one that drops them from the loss (a target floor) and one that raises them to a fixed value of $-10$ (a target clip).
Both are used in \Cref{sec:cost} to show what the treatment of deep cells does to the deployed windows.
The network trained on the raw targets is called the raw-target network below,
and its numbers are the ones reported in the abstract and conclusions.

\subsection{Deployment as Weight Windows}\label{sec:deploy}

A predicted map is written as an OpenMC weight-window file by exponentiating it,
rescaling so that the largest lower bound is 0.5,
as MAGIC does so that source particles born at weight 1 fall inside the source-cell window,
and setting the upper bound to five times the lower.
Before rescaling, every cell below a clamp value of $10^{-6}$ is raised to it.
The clamp is a deployment safeguard whose necessity is demonstrated in \Cref{sec:cost}:
the run cost of a window set is governed by its deepest cells,
and a network that extrapolates a few cells too deep makes the run unaffordable.
The value $10^{-6}$ is the level below which \Cref{sec:val} shows the reference maps themselves to be noise,
so the clamp removes nothing the network could have learned.
It supplies a constraint the training objective cannot:
the mean-squared error in decades is indifferent to a few deep cells,
while the run cost grows exponentially with their depth,
so minimizing prediction error alone does not produce computationally efficient windows.
The map can be applied on the 20~cm mesh on which it is predicted or flux-averaged over $2\times2$ blocks onto a 40~cm mesh;
both are tested.

\section{Results}\label{sec:results}

\subsection{Held-out Ducts}\label{sec:val}

\begin{figure}[t]
    \centering
    \includegraphics[width=0.95\textwidth]{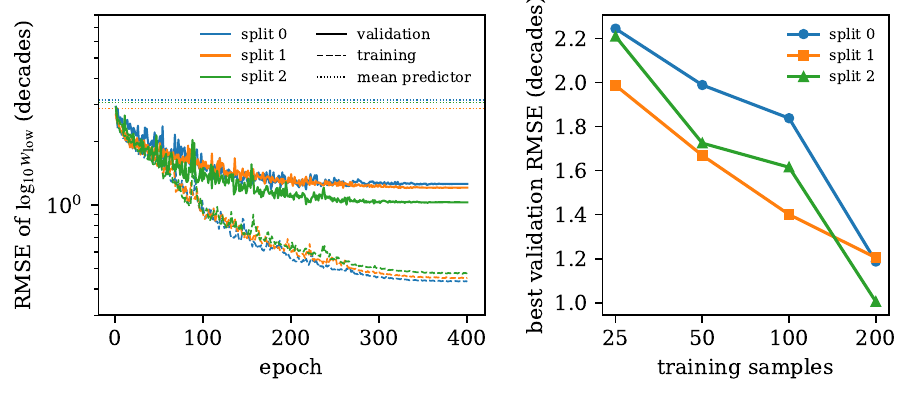}
    \caption{Left: training (dashed) and validation (solid) RMSE against epoch for the three splits;
    dotted lines are the mean-predictor baselines.
    Right: best validation RMSE against the number of training samples for the three splits,
    same \nVal{} validation ducts per split.}
    \label{fig:train}
\end{figure}

\Cref{tab:val} gives the root-mean-square error (RMSE) and mean absolute error (MAE) in decades of $\log_{10} w_{\rm low}$ over all valid cells of the \nVal{} validation ducts of each split,
against a baseline that predicts the training-set mean everywhere.
The pooled RMSE of \unetRmseMin{} to \unetRmseMax{} decades is dominated by the deep tail of the maps,
as \Cref{tab:strat} below shows.
The column ``within $10\times$'' in both tables is the fraction of cells whose predicted $w_{\rm low}$ lies within a factor of ten of MAGIC's,
i.e.\ within one decade.
The left panel of \Cref{fig:train} shows the training and validation curves of the three runs:
validation error stops improving well before 400 epochs while training error continues to fall to \trainRmseMin{} to \trainRmseMax{} decades,
so a generalization gap remains.
The right panel is the learning curve, the best validation RMSE against the number of training geometries.
Retraining with the first 25, 50,
and 100 samples of each training split gives mean validation RMSEs of \lcMeanNTwentyFive,
\lcMeanNFifty, and \lcMeanNHundred{} decades against \lcMeanNTwoHundred{} with all \nTrain,
a curve that has not flattened.
\begin{table}[t]
    \centering
    \caption{Validation error on \nVal{} held-out ducts for three random \nTrain/\nVal{} splits,
    over all valid cells.
    RMSE and MAE in decades of $\log_{10} w_{\rm low}$;
    ``within $10\times$'' is the fraction of cells within one decade of MAGIC.}
    \begin{tabular}{lccccc}
        \toprule
        & \multicolumn{2}{c}{mean predictor} & \multicolumn{3}{c}{U-Net} \\
        split & RMSE & within $10\times$ & RMSE & MAE & within $10\times$ \\
        \midrule
        0 & \baseRmseA & \baseTenA\% & \unetRmseA & \unetMaeA & \unetTenA\% \\
        1 & \baseRmseB & \baseTenB\% & \unetRmseB & \unetMaeB & \unetTenB\% \\
        2 & \baseRmseC & \baseTenC\% & \unetRmseC & \unetMaeC & \unetTenC\% \\
        \bottomrule
    \end{tabular}
    \label{tab:val}
\end{table}

\begin{figure}[t]
    \centering
    \includegraphics[width=0.95\textwidth]{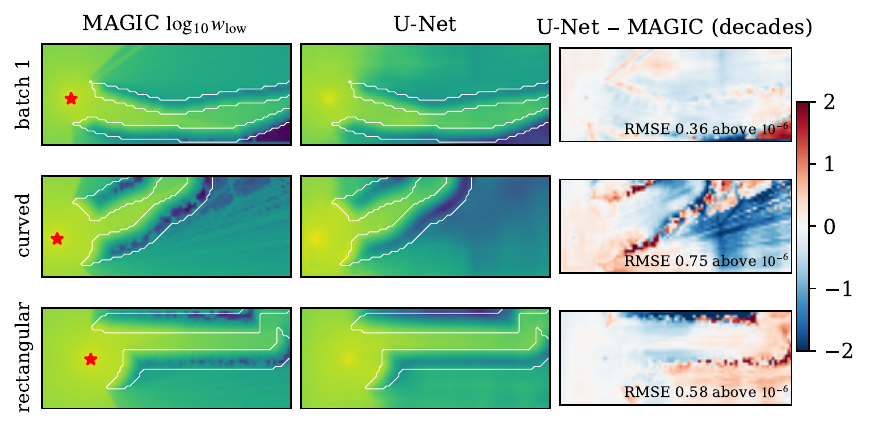}
    \caption{One held-out duct from each family (split~0), drawn with the long axis horizontal.
    Left: MAGIC $\log_{10} w_{\rm low}$ with the concrete outline (white) and source (star); center:
    U-Net prediction on the same color scale; right: signed error in decades,
    with its RMSE over the cells above $10^{-6}$.}
    \label{fig:val}
\end{figure}

\Cref{fig:val} shows one held-out duct of each family.
The network reproduces the source peak, the bright channel, and the drop across each wall;
the error concentrates in the deep shadow,
where the MAGIC map itself shows streak-like scores from a few surviving tracks.
\Cref{tab:strat} quantifies this.
Over the \stratShareBright\% of cells with $w_{\rm low} > 10^{-3}$ the RMSE is \stratRmseB{} to \stratRmseC{} decades with \stratTenC{} to \stratTenA\% of cells within a factor of ten,
and it rises monotonically with depth to \stratRmseF{} decades in the \stratShareF\% of cells below $10^{-6}$.
The RMS deviation of each MAGIC cell from the mean of its same-material $3\times3$ neighbors,
a proxy for the noise of the reference,
grows in the same bins from \roughB{} to \roughF{} decades (last column).

\begin{table}[t]
    \centering
    \caption{Validation error by MAGIC importance level, pooled over the three splits (\stratPixels{} cells),
    and the roughness proxy of the MAGIC target in the same bins (RMS deviation from same-material $3\times3$ neighbors).}
    \begin{tabular}{lcccc}
        \toprule
        MAGIC $\log_{10} w_{\rm low}$ & share of cells & U-Net RMSE & within $10\times$ & MAGIC roughness \\
        \midrule
        $(-1,-0.3]$ & \stratShareA\% & \stratRmseA & \stratTenA\% & \roughA \\
        $(-2,-1]$ & \stratShareB\% & \stratRmseB & \stratTenB\% & \roughB \\
        $(-3,-2]$ & \stratShareC\% & \stratRmseC & \stratTenC\% & \roughC \\
        $(-4,-3]$ & \stratShareD\% & \stratRmseD & \stratTenD\% & \roughD \\
        $(-6,-4]$ & \stratShareE\% & \stratRmseE & \stratTenE\% & \roughE \\
        below $-6$ & \stratShareF\% & \stratRmseF & \stratTenF\% & \roughF \\
        \bottomrule
    \end{tabular}
    \label{tab:strat}
\end{table}

A network fitted by mean-squared error learns the expected map, not the streaks of one run,
so its error has a floor set by the noise of the target.
To measure that floor, three first-family geometries spanning the depth range of the set were rerun with two further random seeds at the production statistics,
and $\sigma_{\rm noise} = \mathrm{RMSE}(\text{replicate}_a, \text{replicate}_b)/\sqrt{2}$ was formed for each of the three pairs of seeds and averaged.
Over the whole map $\sigma_{\rm noise}$ is \nfShallowAll{} decades for the shallowest of the three,
\nfMidAll{} for the intermediate one, and \nfDeepAll{} for the deepest, whose target reaches $10^{-11}$.
Within the deepest geometry, binned as in \Cref{tab:strat}, it is at most \nfDeepBright{} decades for $w_{\rm low} > 10^{-3}$,
\nfDeepMinusFour{} in $(10^{-4}, 10^{-3}]$, \nfDeepMinusSix{} in $(10^{-6}, 10^{-4}]$, and \nfDeepTail{} below $10^{-6}$.
The network error in the same bins, \stratRmseB{} to \stratRmseD{} decades above $10^{-4}$ and \stratRmseE{} in $(10^{-6}, 10^{-4}]$,
is ten to a hundred times larger than the noise there,
so it cannot be attributed to the target and is model error, which more training data reduce, as the learning curve confirms.
Below $10^{-6}$ the target noise is of the order of a decade,
and only more particles per geometry, or averaged replicates, can remove it; more geometries cannot.
The three replicated geometries are from the first family, whose maps end near $10^{-11}$;
the second and third families reach $10^{\targetMinMin}$, and their tail noise, not measured, is larger still.

One systematic error is worth isolating because it interacts with deployment.
Over the \peakBiasMaps{} validation maps of the three splits, \peakBiasDucts{} distinct ducts, the network under-predicts the source cell,
the single sharp peak of every map, in every case, by \peakBiasMin{} to \peakBiasMax{} decades and \peakBias{} on average,
in curved and rectangular families alike,
while a fit of prediction against target over all validation cells above $10^{-6}$ has a slope of \fitSlope{} and an intercept of \fitIntercept{} decades,
so elsewhere it is unbiased.
Rescaling a deployed file to a maximum of 0.5 therefore lifts the rest of the map by about half a decade.
\Cref{sec:cost} shows that a uniform change of level does not affect the figure of merit,
so this bias costs nothing in efficiency,
but it does mean that deployed values must be compared with MAGIC's as gradients rather than levels.

\subsection{Transfer to the Protvino Labyrinth}\label{sec:caseA}

\begin{figure}[t]
    \centering
    \includegraphics[width=0.6\textwidth]{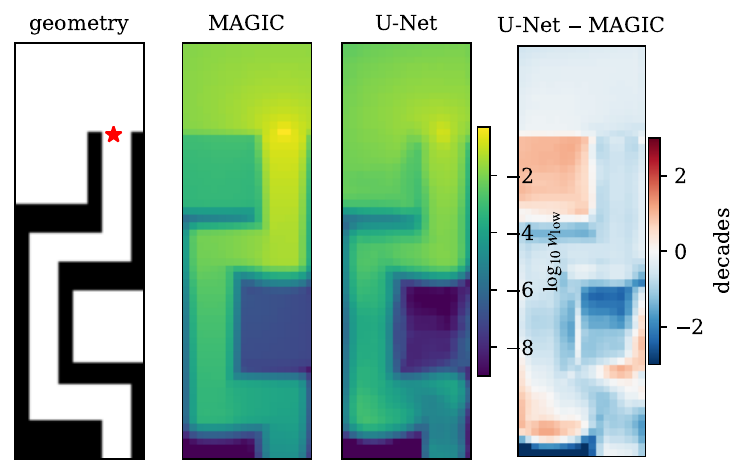}
    \caption{Extended Protvino labyrinth: geometry with source, MAGIC $\log_{10} w_{\rm low}$ at 40~cm,
    U-Net prediction flux-averaged to 40~cm, and their difference.
    The color scale is truncated at $10^{-9}$ so that the corridor values remain distinguishable;
    deeper cells saturate.}
    \label{fig:caseA}
\end{figure}

As a transfer test,
the three trained networks were applied without retraining, as an equally weighted ensemble of independently trained members,
to the extended Protvino labyrinth (case~A of Ref.~\cite{Miao2025MCVR}):
nine rectangular concrete blocks in a $720\times2320$~cm box,
converted without approximation to the 20~cm grid ($116\times36$ cells, every block edge on a cell boundary),
with the source at the origin.
The labyrinth shares the materials, source, and, with the third family, the rectangular construction of the training set,
so the test measures transfer to a new geometric configuration,
not generalization across transport conditions.
The ensemble triples the training cost, about ten minutes per member in parallel,
and predicts in \predMsEnsemble{} milliseconds against \predMsSingle{} for a single network.
The reference is a MAGIC run of 100,000 particles for 1000 batches on an $18\times58$ mesh at 40~cm,
so the prediction is flux-averaged to 40~cm.
\Cref{fig:caseA} shows the result.
Over the \caseACells{} cells the correlation is \caseACorr, the mean offset $\caseAOffset$ decades,
and \caseATen\% of cells lie within a factor of ten of MAGIC;
over the cells above $10^{-6}$ the RMSE is \caseARmseBright{} decades with \caseATenBright\% within a factor of ten.
The spread of the ensemble, the standard deviation of the three members' predictions per cell,
has a median of \caseAEnsSdBright{} decades over the cells above $10^{-6}$ and \caseAEnsSdDeep{} below:
the members agree where \Cref{sec:val} shows the reference to be learnable and diverge where it is noise,
a predictive-uncertainty measure distinct from the Monte Carlo noise of the reference itself.
The rectangular family in the training set has removed the block-interior error of a network trained on curved ducts alone.
The remaining error is in the \caseANDeep{} cells below $10^{-6}$,
where the network extrapolates to a median of $10^{\caseADeepMedianCnn}$ against MAGIC's $10^{\caseADeepMedianMagic}$ and to a minimum of $10^{\caseAMinCnn}$ against $10^{\caseAMinMagic}$;
the training maps reach $10^{\targetMinMin}$,
and the network extrapolates that depth onto a geometry whose corridors do not reach it.

\subsection{Figure of Merit with the Predicted Windows}\label{sec:fom}

The predicted map was written as a weight-window file as described in \Cref{sec:deploy} and used in fixed-source runs of the labyrinth with 100,000 particles per batch for 1000 batches,
one process with 24 threads, with statepoints every 10 batches.
Five reference runs share this configuration: analog; the MAGIC generation run itself,
which builds its windows on the fly and is what a new geometry would have to pay for;
the same generation with the 1000-split cap of \Cref{sec:data};
a run reusing the finished MAGIC file; and that reuse run capped.
With the exit-detector tally of Ref.~\cite{Miao2025MCVR}, its relative error (RE),
and the figure of merit $\mathrm{FOM} = 1/(T\,\mathrm{RE}^2)$ with $T$ the active wall time,
\Cref{tab:fom} compares the runs at 1000 batches.
No training or data-generation time is charged to the CNN,
and none of the runs is capped unless the table says so.

\begin{table}[t]
    \centering
    \caption{Exit-detector figure of merit on the extended Protvino labyrinth at 1000 batches of 100,000 particles,
    one process with 24 threads.
    ``Floor-trained'' and ``clamped'' refer to the networks of \Cref{sec:net} and the deployment of \Cref{sec:deploy};
    the clamped rows are the configuration reported in the abstract.}
    \begin{tabular}{llccccc}
        \toprule
        windows & mesh & RE & s per batch & hours & FOM & vs.\ analog \\
        \midrule
        none (analog) & -- & \fomAnRe & \fomAnSpb & \fomAnHours & \fomAn & 1 \\
        MAGIC, generation run & 40 cm & \fomGenRe & \fomGenSpb & \fomGenHours & \fomGen & \fomGenRatio \\
        MAGIC, generation run, capped & 40 cm & \fomGenCapRe & \fomGenCapSpb & \fomGenCapHours & \fomGenCap & \fomGenCapRatio \\
        MAGIC, file reused & 40 cm & \fomMgRe & \fomMgSpb & \fomMgHours & \fomMg & \fomMgRatio \\
        MAGIC, file reused, capped & 40 cm & \fomMgCapRe & \fomMgCapSpb & \fomMgCapHours & \fomMgCap & \fomMgCapRatio \\
        CNN, floor-trained & 40 cm & \fomFlFRe & \fomFlFSpb & \fomFlFHours & \fomFlF & \fomFlFRatio \\
        CNN, floor-trained & 20 cm & \fomFlTRe & \fomFlTSpb & \fomFlTHours & \fomFlT & \fomFlTRatio \\
        CNN, clamped at deployment & 40 cm & \fomClFRe & \fomClFSpb & \fomClFHours & \fomClF & \fomClFRatio \\
        CNN, clamped at deployment & 20 cm & \fomClTRe & \fomClTSpb & \fomClTHours & \fomClT & \fomClTRatio \\
        \bottomrule
    \end{tabular}
    \label{tab:fom}
\end{table}

The clamped predicted windows on the 20~cm mesh reach a relative error \clampReOverMagic{} times that of the reused MAGIC file at \clampSpbOverMagic{} times its cost per batch,
for a figure of merit of \fomClT{} against \fomMg, or \clampOverReuse{} of the MAGIC file's.
That comparison measures the quality of the predicted map,
since the reused file is a finished product, but it is not the situation a new geometry presents.
A new geometry has no file, and the MAGIC route is the generation run: \fomGen,
or \fomGenRatio{} times analog,
because that run spends its early batches close to analog while it builds the windows and then applies them on the fly.
The predicted windows give \clampOverGen{} times its figure of merit and, in \Cref{tab:ttt},
reach a detector relative error of 0.1\% in \ttTenthClT{} against \ttTenthGen{} for the generation run and \ttTenthMg{} for the reused file,
so for one new geometry they are the fastest route to a given precision among those tested,
and within a factor of \ttTenthClampOverReuse{} of a file that already exists.
Three qualifications apply.
The generation cost depends on its configuration,
and a generation with fewer particles or batches would narrow the difference at some cost in the quality of the windows it produces.
A single geometry is being compared.
And each row of \Cref{tab:fom} is one run with one seed,
so the figure-of-merit values carry the unquantified variability of a single calculation;
repeated runs with independent seeds are left for the full study.

\begin{table}[t]
    \centering
    \caption{Wall time to bring the exit-detector relative error below a target on the extended Protvino labyrinth,
    read from the statepoints of \Cref{tab:fom}.
    Values marked $^*$ are extrapolated from the last statepoint with $\mathrm{RE} \propto T^{-1/2}$.
    The last column adds a capped generation run to a run that reuses its windows.}
    \small
    \begin{tabular}{lcccccc}
        \toprule
        target RE & analog & MAGIC gen. & MAGIC file & CNN 40 cm & CNN 20 cm & capped gen.\ + file \\
        \midrule
        1\%   & \ttOneAn   & \ttOneGen   & \ttOneMg   & \ttOneClF   & \ttOneClT   & \ttOneCapRoute \\
        0.5\% & \ttHalfAn  & \ttHalfGen  & \ttHalfMg  & \ttHalfClF  & \ttHalfClT  & \ttHalfCapRoute \\
        0.2\% & \ttFifthAn & \ttFifthGen & \ttFifthMg & \ttFifthClF & \ttFifthClT & \ttFifthCapRoute \\
        0.1\% & \ttTenthAn & \ttTenthGen & \ttTenthMg & \ttTenthClF & \ttTenthClT & \ttTenthCapRoute \\
        \bottomrule
    \end{tabular}
    \label{tab:ttt}
\end{table}

The capped rows answer the objection that the cap of \Cref{sec:data} might make generation cheap enough to change this ordering.
Capping the generation run produces the same windows in \capGenSpeed{} times less time,
but the run's own detector error is \capGenReRatio{} times worse and its figure of merit,
\fomGenCap, is below analog.
Capping the reuse run, which starts from a finished map, is worse still:
the detector error degrades by a factor of \capReuseReRatio{} for a saving of \capReuseCostSaving\% in cost,
and the figure of merit falls from \fomMg{} to \fomMgCap.
The cap removes exactly the splitting that carries particles down the corridor to the detector,
while the mesh flux that drives the window update, an average over 1044 cells, survives.
So the cap is appropriate for a run whose only product is a window map, as every training run is,
and not for one that must produce the answer;
the last column of \Cref{tab:ttt} shows that generating capped and then reusing the windows costs more than generating uncapped,
and a capped generation used for the answer would need \ttTenthCapGenHours{} hours to reach 0.1\%.

\subsection{What Sets the Run Cost of a Window Set}\label{sec:cost}

The floor-trained network in \Cref{tab:fom} is instructive:
its windows are \floorSpbOverMagic{} times as expensive per batch as MAGIC's file but reach a relative error \floorReOverMagic{} times worse,
and the clamped network on the same mesh buys back that error at \clampFSpbOverFloor{} times the cost for a net gain of \clampOverFloor{} in figure of merit.
Both effects, and the failure of two further window sets that did not finish, have one explanation.
Weight windows hold a particle in cell $i$ at a weight of about $w_i$, the lower bound there,
while the total weight flowing through the cell is the physical flux $\phi_i$ that every unbiased run estimates.
The number of tracks needed to carry it is therefore $\phi_i/w_i$,
and the run cost scales as $\sum_i \phi_i/w_i$.
For MAGIC's own file $w_i \propto \phi_i$, so every cell holds the same population,
which is the purpose of the method.
\Cref{tab:cost} evaluates this proxy for six window sets on the 40~cm mesh with one flux array,
that of the reused-MAGIC run, and compares it with the measured cost per batch.
Proxy and measurement agree within \cpMaxDev\% on all six ratios, which span two orders of magnitude;
the proxy is evaluated from the map before the run,
so this is a prediction of the cost ratio, not a correlation fitted after the fact.

\begin{table}[t]
    \centering
    \caption{Population proxy $\sum_i \phi_i/w_i$ for the run cost of a window set on the 40~cm mesh,
    relative to MAGIC's file, against the measured cost per batch relative to the MAGIC reuse run;
    map minimum in $\log_{10} w_{\rm low}$.
    ``Top cell'' is the share of the proxy in the single most populated cell.
    The unclamped rows are the two alternative networks of \Cref{sec:net} and the raw-target network deployed without the clamp;
    those runs were stopped after 20~h.}
    \small
    \begin{tabular}{lcccc}
        \toprule
        windows & map minimum & proxy cost & measured cost & top cell \\
        \midrule
        MAGIC file & \cpMgMin & \cpMgProxy & \cpMgMeas & \cpMgTop\% \\
        CNN, earlier 50-geometry network & \cpFiftyMin & \cpFiftyProxy & \cpFiftyMeas & \cpFiftyTop\% \\
        CNN, floor-trained & \cpFlMin & \cpFlProxy & \cpFlMeas & \cpFlTop\% \\
        CNN, clip-trained, unclamped & \cpCpMin & \cpCpProxy & \cpCpMeas & \cpCpTop\% \\
        CNN, raw-target network, unclamped & \cpNfMin & \cpNfProxy & \cpNfMeas & \cpNfTop\% \\
        CNN, raw-target network, clamped & \cpClMin & \cpClProxy & \cpClMeas & \cpClTop\% \\
        \bottomrule
    \end{tabular}
    \label{tab:cost}
\end{table}

The table makes three points.
First, the cost of a window set is governed by its deepest cells, not by its accuracy:
the raw-target network deployed without the clamp is \cpNfMeas{} times as expensive as MAGIC's file because in \cpNfDeepCells{} of the 1044 cells it lies more than three decades below MAGIC,
and those cells hold most of the predicted population,
\cpNfTop\% of it in a single dead corner where the network predicted \cpNfTopDecades{} decades deeper than MAGIC.
Its run reached \fomNfFBatches{} batches in 20~hours.
The clip-trained network, which bottoms out at $10^{\cpCpMin}$, still cost \cpCpMeas{} times MAGIC.
Second, the floor-trained network is cheap for the same reason,
since dropping the deep cells from the loss left it unconstrained there and it saturated near $10^{\cpFlMin}$;
but the same saturation brightened the exit leg of the corridor,
so the run rouletted the particles that feed the detector and its statistics were poor.
Third, the clamp separates the two effects:
it caps the deployed depth at the same level as the floor-trained map while keeping the corridor shape the raw-target network learned,
and the proxy predicted its cost, \cpClProxy{} of MAGIC's, before the run.
The treatment of deep cells is therefore a deployment choice, not a training one,
and the clamp value is a knob that trades cost against detector precision.
A uniform change of level, by contrast,
is equivalent to changing the number of particles per batch and leaves the figure of merit unchanged to first order,
which is why the source-cell bias of \Cref{sec:val} costs nothing here.

Two further observations.
The finer 20~cm mesh improves the figure of merit by \meshGainClamp\% for the clamped windows and \meshGainFloor\% for the floor-trained ones,
while it worsened it by a quarter for the 50-geometry network of an earlier stage of this work;
the effect is within the scatter of single runs.
And the corridor-wide view from the mesh flux tally, a figure of merit per cell, does not separate the window sets:
the median over corridor cells is \meshMgCorrMedian{} for MAGIC's file and \meshClTCorrMedian{} to \meshClFCorrMedian{} for the clamped windows,
and the worst corridor cell is set by single cells that a corridor-wide evaluation would have to run longer to resolve.

\section{Conclusions and Next Steps}\label{sec:concl}

On the evidence so far,
a U-Net trained on \nTrain{} random ducts of three families predicts the lower window bound of unseen ducts to \stratRmseB{} to \stratRmseD{} decades in the cells that carry the flux,
from geometry and source position in \predMsSingle{} milliseconds on a CPU,
against \rtUncMeanMin{} minutes of uncapped MAGIC iteration per geometry on average.
Replicate MAGIC runs show that this part of the map is reproducible to a few hundredths of a decade,
so the residual there is model error,
and the learning curve from 25 to \nTrain{} samples has not flattened;
below $10^{-6}$ the reference is noise.
On the rectangular Protvino labyrinth the predicted windows, clamped at deployment,
give an exit-detector figure of merit \fomClTRatio{} times analog,
\clampOverGen{} times that of the MAGIC generation run a new geometry would otherwise require,
and \clampOverReuse{} of that of a finished MAGIC file,
reaching 0.1\% precision in \ttTenthClT{} against \ttTenthGen{} for the generation run.

Two findings about the windows themselves go beyond the network.
The run cost of any window set is governed by its deepest cells,
and the population proxy $\sum_i \phi_i/w_i$, evaluated from the map before the run,
matched the measured cost per batch within \cpMaxDev\% over six window sets spanning two orders of magnitude,
so the cost of a map can be estimated before it is run and an extrapolating network must be clamped at deployment.
A cap on splits per history makes the generation of training maps \ccSpeedMin{} to \ccSpeedMax{} times cheaper without changing the maps above $10^{-6}$,
but destroys the figure of merit of any run that must also produce the answer,
which separates the two uses of a Monte Carlo run cleanly.

The training set cost \rtNodeHours{} node-hours,
equivalent to \nodeHoursOverGen{} uncapped generation runs of the benchmark,
and a capped one costs \rtCapMeanMin{} minutes on one node, so the dataset can be enlarged cheaply.
The next steps are to (i) continue the learning curve to a thousand geometries,
since it has not flattened and the flux-carrying region is not noise-limited; (ii) tune the clamp,
since a clamp one decade deeper preserves the exit leg as predicted at a projected 15\% more cost;
(iii) vary the pixel size; and
(iv) test the predicted map as a warm start for the MAGIC iteration,
which loses its early batches converging from analog windows.

\section*{AI USAGE STATEMENT}
Portions of this work,
including code development, plotting, and the linguistic refinement and structural editing of the manuscript,
were assisted by artificial-intelligence tools under the authors' direction.
All content was reviewed and verified by the authors.
The authors take full responsibility for all content of this work,
and AI usage cannot be justified as a reason for any errors that might be found in this work in the future.

\section*{ACKNOWLEDGEMENTS}
The authors thank the Penn State Department of Nuclear Engineering for support and the Institute for Computational and Data Sciences for HPC resources.

\bibliographystyle{ANStopical}
\bibliography{references}

\end{document}